# Coupling in Quantum Dot Molecular Hetero-Assemblies

Carlo Nazareno Dibenedetto[a,b], Elisabetta Fanizza[a,b], Liberato De Caro[c], Rosaria Brescia[d], Annamaria Panniello[b], Raffaele Tommasi[e,b], Chiara Ingrosso[b], Cinzia Giannini[c], Angela Agostiano[a,b], Maria Lucia Curri[a,b], Marinella Striccoli[b,*]

[a.] Department of Chemistry, University of Bari ''Aldo Moro'', Via Orabona, 4 - 70125 Bari (Italy).

[b.] Institute for Chemical and Physical Processes of CNR (IPCF-CNR), Via Orabona, 4 - 70125 Bari (Italy). *corresponding author: M.S. m.striccoli@ba.ipcf.cnr.it;

[c.] Institute of Crystallography of CNR (CNR-IC), Via Amendola, 122/O - 70125 Bari (Italy).

[d.] Italian Institute of Technology (IIT), Via Morego 30 - 16163 Genova (Italy).

[e.] Basic Medical Sciences, Neuroscience, and Sense Organs, University of Bari ''Aldo Moro'', Piazza Giulio Cesare, 11 - 70124 Bari (Italy).

Abstract: The design of large-scale colloidal quantum dots (QDs) assemblies and the investigation of their interaction with their close environment are of great interest for improving QD-based optoelectronic devices' performances. Understanding the interaction mechanisms taking place when only a few QDs are assembled at a short interparticle distance is relevant to better promote the charge or energy transfer processes. Here, small hetero-assemblies formed of a few CdSe QDs of two different sizes, connected by alkyl dithiols, are fabricated in solution. The interparticle distance is tuned by varying the linear alkyl chain length of the bifunctional spacer from nanometer to sub-nanometer range. The crystallographic analysis highlights that the nearest surfaces involved in the linkage between the QDs are the (101) faces. The thorough spectroscopic investigation enables a sound rationalization of the coupling mechanism between the interacting nanoparticles, ranging from charge transfer/wavefunction delocalization to energy transfer, depending on their separation distance.

Highlights:

- Molecular assemblies of Quantum Dots (QD) are prepared in solution, linked by dithiols.
- The interparticle distances are tuned by varying the linear alkyl chain length of the dithiols from nanometer to sub-nanometer range.
- The crystallographic analysis highlights that the (101) faces of the QDs play an important role in the linkage.
- The hetero-assemblies are spectroscopically investigated in solution and compared with the homolog homo-assemblies.
- QD coupling is governed by the interparticle distance, the QD size, and the mutual energetic levels arrangement.

## 1. Introduction

In the last few years, a great scientific effort has been made to deeply understand the coupling mechanisms among colloidal quantum dots (QDs) interacting with each other when spaced at nanometric and sub-nanometric distances.[1-9] The elucidation of the interaction mechanisms among the QDs both in solution and in solid-state can open the opportunity to realize highly efficient charge transfer systems for optoelectronic devices and take a jump towards new fancy applications like quantum computing,[2, 3, 10] spintronic devices,[11] biosensors,[12, 13] solar cells[14, 15] and chiral-induced spin selectivity effect.[16] In these applications, the surface chemistry of the QDs, often controlled by the ligands,[17] plays a key role and an adequate understanding of the type of coupling involved that takes also in consideration the role of the organic capping layer is needed. Classically, when nanoparticles interact, two main mechanisms can be distinguished, namely a pure charge transfer that can be modeled by Dexter's theory,[18] and an energy transfer induced by the dipole-dipole coupling (FRET), explained by the Förster theory.[19, 20] Both processes can be depicted as an electron or dipole moment transfer, respectively, from QDs acting as donors (D) to QDs acting as acceptors (A). In particular, these interaction phenomena are very sensitive to the distance among the nanoparticles that can range from sub-nanometer to tens of nanometers.

At very short distances (i.e. less than ~1 nm) the interaction leads to the overlap of the electronic wavefunctions of the nano-objects, resulting in a charge transfer (contact zone, Dexter model), while for longer distances (from ~1 nm up to ~10 nm) the dipole-dipole coupling (near field zone, Förster theory) plays the major role, resulting in FRET.

Many reports[21-25] have tried to distinguish the different contributions to the transfer efficiency made by FRET and non-FRET processes, mainly investigating the interaction between dye molecules,[26, 27] and between nanocrystals and dye molecules. In particular, Moroz et al.[21] studied the FRET/non-FRET contributions in cyanine dye–QD solid-state assemblies by a spectroscopic approach to calibrate a spectroscopic ruler. Hoffman et al.[22] highlighted how a nanostructured system, composed of CdSe QDs of different sizes, linked to a red-infrared-absorbing squaraine dye through a thiol functional group, couples via energy transfer through both long-range dipole-based FRET and short-range Dexter electron transfer mechanisms. Avellini et al.[23] and Harris et al.[24] reviewed the charge/energy transfer in QDs/molecules assemblies reporting the possible interactions between the chromophores when in close proximity, either connected chemically by ligand or layered in disordered solid films. Panniello et al.[25] developed an efficient FRET system combining QD donors and dye molecules acceptors, namely BODIPY, and studied how different D-A distances can provide effective energy transfer already in solution, with an efficiency of 76% that increases in solid-state. Few studies report, so far, QD-QD coupling. Kolodny et al.[6] demonstrated that the linkers used to connect the QDs can modify their coupling through vibration-assisted transport. Zheng et al.[28] investigated FRET dynamics in densely packed films composed of multi-sized CdSe QDs using ultrafast transient absorption spectroscopy and theoretical modelling, being finally able to separate FRET contribution from intrinsic exciton decay. Hoffman et al.[29] explored the impact of QD surface chemistry on energy transfer in films formed by differently sized CdSe QDs, spin-casted onto a glass substrate. A solid-state ligand exchange with thiol was implemented to investigate the effects of QDs surface passivation on energy transfer and to mimic a layer-by-layer deposition commonly used in the fabrication of QD photovoltaic devices. Kim et al.[30] proposed a 3D CdSe QD superlattice and demonstrated a quantum resonance among the QDs decreasing their size and the distance between the QD layers, rather than a dipole-dipole Coulomb coupling. Similarly, Sugimoto et al.[31] performed a detailed spectroscopic studies on silicon QDs assemblies tuning the distance among the nano-objects and their size distribution.

Most of these studies use disordered solid-state films of QDs, assembled or spin-coated on a substrate, where the aggregation of QDs and their collective interactions can play a predominant role, thus somehow masking the fundamental interaction occurring ideally between two single QDs at a given distance. To overcome such a limitation that arises in the solid-state systems, we previously demonstrated the fabrication of a few linked equal-sized QDs in the solution, namely homodimers.[5] Such nanostructures couple through wavefunction delocalization at a sub-nanometric interparticle distance, while a FRET behaviour is assessed for longer distances. Recently, also Cui et al.[7] obtained homodimers of CdSe/CdS core-shell nanocrystals by constrained oriented attachment and their coupling was proven by the red-shift of the excitonic transitions.

However, such homodimer systems present limitations due to the difficult discrimination between donor and acceptor QDs. The realization of QDs hetero systems, in solution, could overcome this issue, giving a better comprehension of the coupling phenomena involved.

Here, molecular hetero-assemblies have been fabricated starting from CdSe QDs of two different sizes, synthesized to have the first excitonic transition of the donor in mutual resonance with the second excitonic transition of the acceptor. The QDs are linked by linear dithiols (DT) of different alkyl chain lengths in the nanometric and sub-nanometric regimes. The thiol group ensures the chemical binding with the QDs,[32] forming with the cadmium on the QDs surface an X-type bond.[33] We performed an in-depth morphological and spectroscopic characterization to get information about the crystallographic faces involved in the bound and in the nature of the coupling between the nano-objects. Steady-state and time-resolved photoluminescence (PL and TR-PL, respectively) measurements on the hetero assembled system and the comparison with similar experiments carried out on homo-assemblies allowed to estimate the spacing range between the nanoparticles in which the charge transfer is the dominant mechanism. We demonstrated that, for short ligand chains, the wavefunction delocalization in hetero-assemblies is more efficient than in homo ones, while FRET processes dominate in homo-assemblies at larger distances. These findings can be profitably used in the optoelectronic applications that take advantage of cascade-like energy level systems to more efficiently funnel charges with increased transfer rates and be of inspiration in the design of systems for photoconversion, light-harvesting, and catalysis.

2. Experimental

2.1. Materials

Cadmium oxide (CdO, 99.5%), selenium (Se, 99.99%), tributylphosphine (TBP, 99%), trioctylphosphinoxide (TOPO, 99%), hexadecylamine (HDA, 90%), oleic acid (OLEA, 90%), butylamine (BUA, 99%), 1,3 propanedithiol (pDT, 99%), 1,6 hexanedithiol (hDT, 96%), 1,8 octanedithiol (oDT, 97%), 1 propanethiol (pT, 99%). All chemicals were used as received, without any further purification or distillation. Ethanol (≥ 99,8%), hexane (anhydrous ≥ 99%), cyclohexane ($cC_6$, Spectroscopic Grade ≥ 99.9 %) and tetrachloroethylene (TCE, Spectroscopic Grade ≥ 99.9 %) were used at analytical grade, unless otherwise specified. All the chemicals were purchased from Sigma-Aldrich.

## 2.2. Synthesis of CdSe QDs

Following a procedure reported in literature[5] CdSe QDs were synthesized by a colloidal approach. 0.127 g of CdO (1 mmol) were decomposed at 260°C under nitrogen flux by means of 1 mL of OLEA and then cooled to 85 °C. In a second flask TOPO (9 g, 23 mmol) and HDA (9 g, 37 mmol) were heated at 110°C, then cooled to 85 °C and injected into the flask with the Cd precursor. At 280 °C, 2 mL of TBP were injected followed by the injection at 295 °C of the Se precursor (0.394 g, 5 mmol, dispersed in 4.5 mL of TBP). The growth temperature was set at 270 °C. Two samples of QDs were synthesized varying the growing time: at 90 sec (QD1) and at 180 sec (QD2). The QDs were collected by centrifugation, adding ethanol and dispersing the precipitate in 4 mL of hexane, thus resulting in a concentration of $2 \cdot 10^{-4}$ M for QD1 and $1 \cdot 10^{-4}$ M for QD2, respectively, as calculated by.[34]

## 2.3. QD functionalization with butylamine, alkyl thiols and dithiols

Molecular hetero-assemblies were fabricated following a previously reported procedure.[5, 6] First, QDs ($5 \cdot 10^{-7}$ M) were treated by using BUA ($2.5 \cdot 10^{-2}$ M in hexane) at BUA: QD molar ratio of 500:1. The addition of the dithiols (DT, $1.7 \cdot 10^{-4}$ M in hexane) was further carried out at DT:QD molar ratio of 50:1, under stirring for 5 min. To obtain a reference sample, QDs were functionalized with pT. The treatment of BUA functionalized QD1 and QD2 samples ($5 \cdot 10^{-7}$ M) was performed by using pT ($3.3 \cdot 10^{-3}$ M in hexane) in a QD: thiol molar ratio 1:50 for 5 min under stirring at room temperature. The enrichment in molecular assembly of the solution and the purification of the same from the organic impurities and larger aggregates was carried out by means of density gradient ultracentrifugation (DGU) using a Beckman Coulter Ultracentrifuge mod. Optima XE 90. The gradient was composed by 6 layers of 500 μL each of a mix of $cC_6$ and TCE, from 90% to 30% of $cC_{60}$ and built into a thick wall polyallomer round bottom centrifuge tube. The samples were centrifuged at 25000 rpm for 15 min. The layer at 40% in $cC_{60}$ containing the molecular hetero-assemblies was carefully recovered with a syringe and transferred into a glass vial. Spectroscopic characterization such as UV-Vis absorption, PL, and TR-PL ($\lambda_{ex}$=485 nm), together with morphological analysis were carried out on the selected fraction. Samples were diluted to 1:60 in hexane for TEM grid preparation.

## 2.4. Characterization techniques

UV-Vis absorption spectra were recorded with a Cary 5000 (Varian) UV/Vis/NIR spectrophotometer. All fluorescence measurements were performed at room temperature. Fluorescence spectra were acquired by using a Fluorolog 3 spectrofluorometer (HORIBA Jobin-Yvon), equipped with double grating excitation and emission monochromators. TR-PL measurements were performed by Time-Correlated Single Photon Counting (TCSPC) technique, with a FluoroHub (HORIBA Jobin-Yvon). The samples were excited at 485 nm using a picosecond laser diode (NanoLED 485L) with a pulse length of 80 ps at a 1 MHz repetition rate, with typical energy below 10 pJ/pulse. The PL signals were dispersed by a double-grating monochromator and detected by a picosecond photon counter (TBX ps Photon Detection Module, HORIBA Jobin-Yvon). The temporal resolution of the experimental setup was ~200 ps. DAS6 Analysis® software by Horiba was used to fit the decay profiles with multiexponential functions (of order 3 or 4) and average lifetimes[35] in which the QDs remain in the excited state following excitation were calculated as reported in the electronic supplementary information (ESI). Absolute quantum yield measurements were obtained utilizing a ''Quanta-phi'' integrating sphere coated with Spectralons® and mounted in the optical path of the spectrofluorometer, using as excitation source a 450 W xenon lamp coupled with a double-grating monochromator. For transmission electron microscopy (TEM) analysis, samples were prepared by casting 1.5 µL of QDs solution on a carbon-coated copper grid. A JEOL JEM1011 microscope, operating at an accelerating voltage of 100 kV and equipped with a W electron source and a CCD high-resolution camera was used for image acquisition at low resolution. Statistical analysis of the QD average size and size distribution of the samples was performed by using an image analysis software (AxioVision®). The percentage relative standard deviation (σ%) was calculated for each sample, providing information on the QD size distribution. Its value is based on the distribution of size compared to the average value and is expressed as a percentage. The high resolution TEM (HR-TEM) images were acquired by an image-Cs-corrected JEOL JEM-2200FS TEM (Schottky emitter source, operated at 200 kV), with an in-column filter (Ω-type). For the specimen preparation, a small volume of each sample was deposited onto an ultrathin carbon/holey-carbon film-coated Cu grid. The TEM grids, after deposition of the samples, were heated up in the high vacuum, up to 150 °C, to enhance desorption of residues of organics/solvents. The images were frequency-filtered[36] to minimize the contrast due to amorphous carbon support. The fast Fourier transforms of HR-TEM images were compared with single-crystal electron diffraction patterns from two crystal structures: ICSD 41528 for zinc-blende CdSe and ICSD 415785 for wurtzite CdSe. For all HR-TEM images, this procedure provided a better match with the wurtzite CdSe phase. The zone axis provided for each particle in the HR-TEM images was chosen due to the best overall match with the calculated single-crystal diffraction pattern. The slight discrepancy between experimental and calculated inter-plane distances and angles can be explained based on the inherent accuracy of magnification calibration in HR-TEM (affecting the measured interplanar spacings), around 5%,[37] and to slight distortions of the crystal lattice (affecting measured angles) in nanometer-sized particles.

3. Results and discussion

For the molecular assembly's fabrication, CdSe QDs have been synthesized in two different sizes by a classical approach based on thermal decomposition of precursors in hot non-coordinating solvents,[38-40] tailoring the size by controlling the reaction times. The synthesized QDs, dispersed in hexane, result mainly capped by TOPO and HDA, the surfactants used in the synthetic procedure. In Figure S1 in ESI the spectroscopic and the morphologic features of the QDs are reported. In particular, the QD1 is characterized by the first excitonic transition at 590 nm with PL peaked at 600 nm, while the QD2 has the first excitonic transition at 617 nm and the PL centred at 625 nm. The TR-PL decay profiles of QD1 and QD2 reported in Figure S2 in ESI, show similar recombination dynamics, not significantly influenced by the QD size. The diameter of the QDs measured by TEM is 3.7 nm (σ%=8%) for the QD1and 4.4 nm (σ%=7%) for the QD2, in agreement with the spectroscopic size determination.[34] The nanocrystal sizes have been suitable designed so that the QDs have their excitonic transitions in mutual resonance; in particular, the first excitonic transition ($1S_e – 1S_{3/2}$, |1⟩) of the smaller QD1 energetically matches the second excitonic transition ($1S_e – 2S_{3/2}$, |2⟩) of the larger QD2, both centred at 590 nm. Considering the possible energy or charge transfer processes that can occur between the two nano-objects when they are in close proximity,[23, 24, 41, 42] we can attribute at the QD1 the role of donor and at the QD2 the role of acceptor, even when they are dispersed in solution.

A protocol for the fabrication of molecular homodimers,[5] consisting of a two steps functionalization procedure of the QDs has been successfully applied here to the hetero-assemblies preparation (Scheme 1).

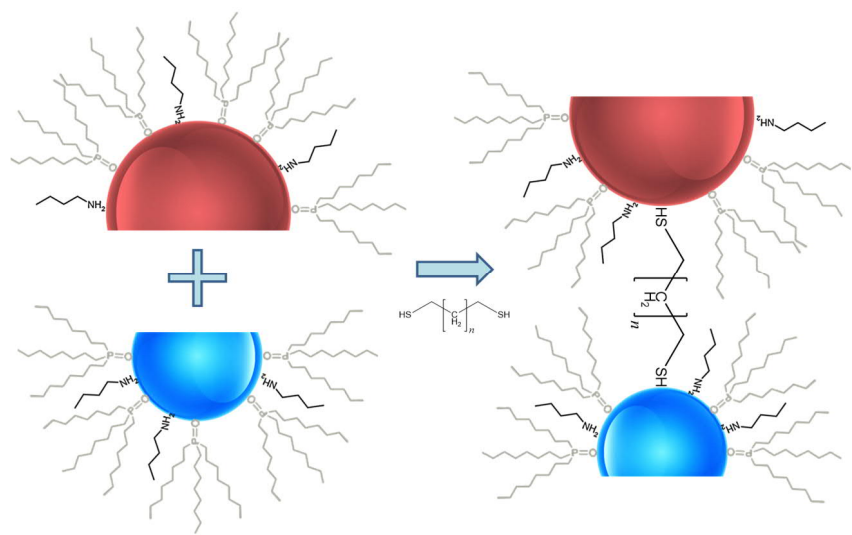

Scheme 1. General strategy for the fabrication of molecular hetero assembly. The anchoring of a bifunctional linker to the surface of pre-functionalized QDs of two different sizes can result in hetero-assembly formation.

Firstly, both QDs have been pre-functionalized with butylamine (BUA), a short alkyl chain ligand, to reduce the large steric hindrance of the native ligands present on the surface of the QD deriving from the synthetic procedure (i.e. mainly TOPO and HDA). Indeed, such a steric obstacle prevents the efficient penetration of the bifunctional linker through the organic shell by precluding it from approaching the QD surface for binding. Since the long alkyl chains of native ligands ensure the colloidal stability of the QD in solution, only a partial replacement is needed to make more accessible the surface of the QDs to the new ligand, still preserving the QD stability and avoiding unwanted and massive aggregation. For this reason, as in ref.[5], a molar ratio BUA:QD of 500:1 has been used, to limit the extent of the ligand exchange. The second step consists of mixing the hexane solutions of the pre-functionalized QD1 and QD2 at the same concentration ($5 \cdot 10^{-7}$ M), followed by the addition of the bifunctional linker suitable to connect two QDs. For the fabrication of molecular hetero-assemblies, bifunctional linkers belonging to dithiol's class, in particular, the 1,3 propanedithiol (pDT), 1,6 hexanedithiol (hDT) and 1,8 octanedithiol (oDT) characterized by a linear alkyl chain length[43] of 0.55 nm, 0.9 nm, and 1.2 nm respectively, have been selected. The dithiols have been added in a molar ratio DT:QD of 50:1 to the pre-functionalized mixed solution of QD1 and QD2 to obtain the highest assembly yield and avoid larger aggregation. In order to purify the sample and enrich the solution in molecular assemblies, separating the unlinked QDs, large aggregates and the organic impurities in solution, a DGU has been performed.[44]

In Fig. 1 the TEM micrographs of the as-prepared small molecular assemblies of QDs (1A) linked with pDT, containing also few unlinked QDs and of the larger molecular assemblies obtained after DGU procedure are reported. The ultracentrifugation results in a purified sample with a visible enrichment in assembled nano-systems composed up to 8 QDs (Fig. 1B). Large aggregates and organic impurities are collected at the bottom layer of the tube and removed (Figure S3). The probability to collect molecular assemblies constituted by the two QDs of different size is very high, which reflects into a high percentage of hetero structures. The assembly statistics follows the hypergeometric distribution that, starting from two different families of objects, A and B, describes the probability to have at least one object A in an ensemble of n objects B.[45] In particular, the probability for an assembly of 4 QDs to have at least one QD of different size is 93.8% and rises up to 99.6% for assemblies of 8 QDs. In Fig. 2 the filtered HR-TEM micrographs of QD1-QD2 hetero-assemblies fabricated with pDT (2A) and hDT (2C) as bifunctional linkers, are shown. The crystallographic analysis of the Fast Fourier Transforms (FFTs) of the HR-TEM images indicates a wurtzite structure (ICSD 415785) of the nanocrystals constituting the assembly, as shown in the enhanced images with respect to the amorphous background reported in Fig. 2B and 2D, respectively.

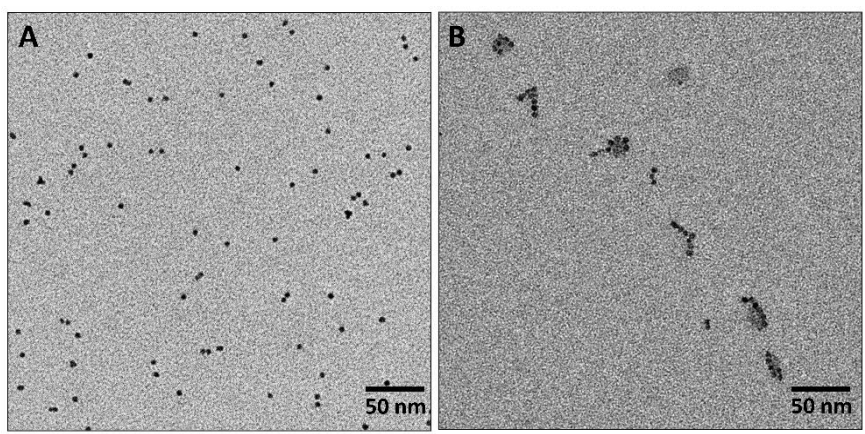

Figure 1. TEM micrographs of the as-prepared molecular assemblies (A) and after DGU procedure, resulting in a visible enrichment of the solution in molecular assemblies (B).

Wurtzite CdSe nanocrystals exhibit a frustrated hexagonal prismatic shape. Panels B and D of Fig. 2 show a 3D-model of the pDT and hDT hetero-assemblies, respectively, that can be deduced by the FFTs' analysis of the HR-TEM images, only when the QDs are aligned along a defined zone axis, as for the examples shown. The zone-axis projection of the 3D-model of the QDs is reported in Figure S4. The crystallographic analysis highlights that the nearest surface regions of the two QDs are the (101) faces. Moreover, the enhanced pDT sample image (Fig. 2B) allows to evidence that the two QDs are so close to each other that their lattice fringes seem to be partially overlapped.

Accordingly, the hDT QDs image in Fig. 2D confirms that the (101) faces are the nearest, although the distance between the two QDs is larger than the corresponding distance for the pDT case.

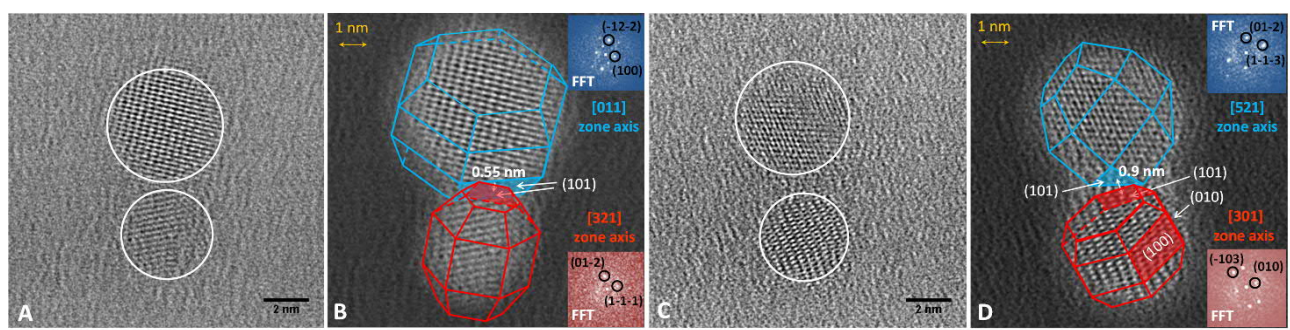

Figure 2. Filtered HR-TEM images of the pDT (A) and hDT (C) hetero-assemblies to minimize high-frequency noise. Further filtering of the HR-TEM images allows to enhance the QDs with respect to the amorphous background (panel B and D). A 3D-model (frustrated hexagonal prismatic shape) of the two QDs constituting the pDT (B) and hDT (D) hetero-assemblies, seen in projection along the respective zone-axis direction is reported. The insets show the FFTs of the two HR-TEM images.

The QDs' distance values that can be estimated by the 3D-model reconstruction (Fig. 2B and 2D) agree with the nominal values of the linear alkyl chain lengths, although a quantitative evaluation for the pDT case is limited by the fact that the 3D QDs are seen in projection and are so close to each other that can be also partially overlapped. In any case, to give a length reference, a segment representing the nominal length of pDT and hDT chains is sketched between the two QDs, rotated towards the image plane of 45° to take into account a reduction of length due to the projection in the image plane. Finally, it should be noted that in CdSe wurtzite: the two terminating (001) faces are cadmium and selenium surfaces, respectively; the (100) faces are mixed cadmium and selenium surfaces; the (101) faces are either cadmium or selenium surfaces.[46] Therefore, from the crystallographic analysis, shown in Fig. 2, the (101) cadmium surfaces seem to have an important role in favouring the assemblies' formation, at least for values of the alkyl chain length up to about 0.9 nm (hDT). However, in the ESI section, a third example is reported (Figure S5) - the crystallographic analysis for an oDT sample, whose chain length is of 1.2 nm - for which the nearest surfaces of the two QDs are (100) and (001), different from the (101) ones found for the pDT and hDT cases.

In the following, a comparison of the optical properties of hetero-assemblies, unlinked QDs and homo-assemblies in solution is reported to underline the different spectroscopic behaviour and assess the transfer mechanism between the nanocrystals.

3.1. Comparison between hetero-assemblies and unlinked QDs in solution

Spectroscopic measurements as absorption, steady-state and TR-PL can provide meaningful information on the interaction between the two differently-sized QDs. It is well-known that the thiol moiety functionalizing the CdSe QD surface induces a quenching of the PL emission because the sulphur generates a trap state for the holes into the bandgap near the valence band,[47] also demonstrated by XPS measurements.[48] To discriminate the quenching effect due to the sole thiol moiety from the additional quenching induced by a possible interaction between the linked QDs, the 1 propanethiol (pT) functionalized QDs have been chosen as reference, as pT can bind the surface of one single QD with the thiol moiety, but cannot induce the particle's connection, since it is not bifunctional. In Fig. 3A the PL spectra of two molecular hetero-assemblies fabricated by using pDT and hDT as ligands (red and green lines, respectively) compared with the simple mixed QDs (dark blue line) and the reference sample with pT (light blue line) are reported.

The thiol moiety bound at the QDs surface induces a quenching of the emission, as can be noticed for all the samples treated with thiol and dithiols in Fig. 3A. Such a quenching is only mild for QDs treated with pT, while appears much intense for dithiols, also depending on the length of the alkyl chain of the ligand. A more pronounced quenching of the QD1 emission with respect to the QD2 in the PL spectra can be also observed, still depending on the alkyl chain length of the bifunctional molecules. Besides, normalized PL spectra (Fig. 3B) show a clear redshift of the emission in these samples with respect to the mixed QDs, caused by a reduction of the high-energy emission and an enhancement of the contribution in the red region.[41] The redshift is observed also in the absorption spectrum of QD1-QD2 pDT reported in Figure S6 in the ESI. Such phenomena suggest the occurrence of an interaction between the QDs when they are closely connected by dithiols, similarly to what has been found for homodimers.[5]

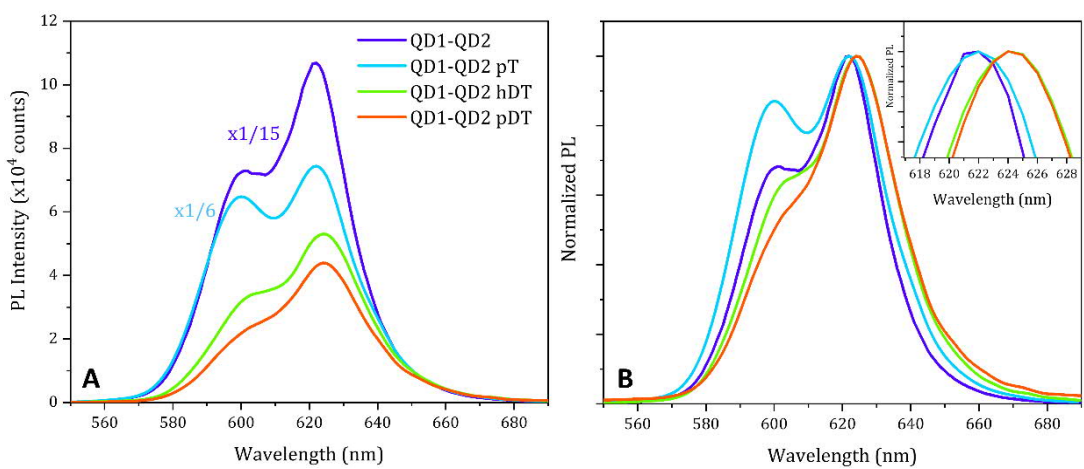

Figure 3. PL spectra of molecular hetero-assemblies obtained using pDT (red line) and hDT (green line) as linkers compared with the only mixed and pT treated QDs (A) and after normalization (B). The values of the intensity of QD1-QD2, mixed (dark blue line) and treated with pT (light blue line) have been divided by 15 and 6, respectively. In the inset of panel (B) a close-up of the peak profiles. $\lambda_{ex}$=485 nm

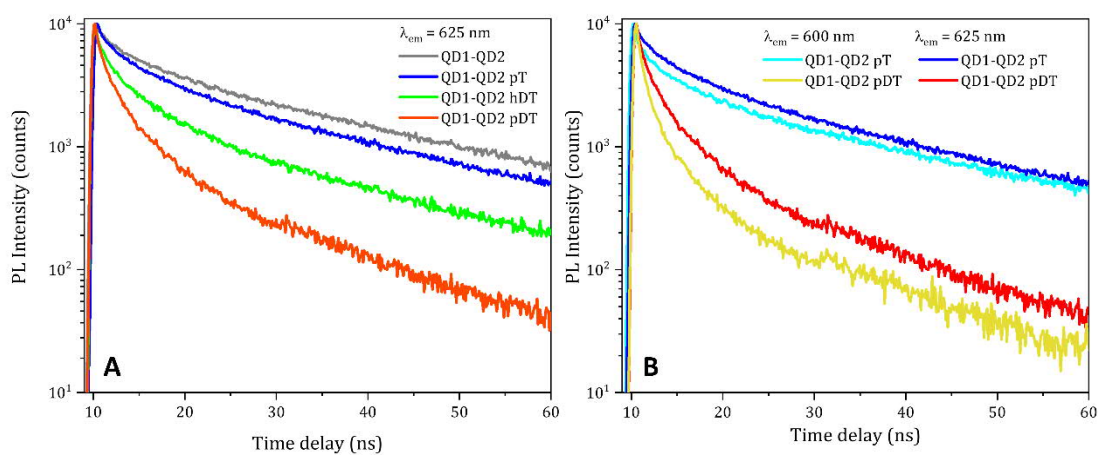

Figure 4. A) TR-PL of hetero-assemblies prepared with pDT (red line) and hDT (green line), compared with the decay profiles of the mixed QD1-QD2 (grey line) and treated with pT (blue line). B) Comparison of the decay profiles of the mixed QD1-QD2 treated with pT (cyan and blue lines) and the pDT hetero-assembly (red and yellow lines), measured in correspondence of the emission peaks of the two QDs. $\lambda_{ex}$=485 nm

Fig. 4A displays the TR-PL decay profiles of the samples prepared with pDT, hDT and of the QD mixture and pT-treated, as a reference, measured in correspondence of the emission peak of QD2. The dynamics have been best fitted with multi-exponential functions and the resulting average lifetimes, reported in table S1 in the ESI section, decrease from 26.0 ns for the mix QDs solution down to 8.6 ns for the QDs functionalized with pDT, while the pT induces only a small reduction in the lifetime (25.1 ns). The molecular assembly's lifetimes decrease when reducing the interparticle distance, in agreement with the quenching of PL (Table S1 in the ESI). In Fig. 4B the comparison of the PL decay profiles of the pDT hetero-assemblies and mixed QDs with pT, collected at the emission wavelength of QD1 (600 nm) and QD2 (625 nm), is reported. While for the QDs treated with pT the difference between the recombination dynamics at the two wavelengths is not so evident and follows the behaviour of the pure QDs in solution (Figure S2 in the ESI), for the hetero-assembly the recombination of QD1 is faster than that of QD2. If, as previously mentioned, QD1 is considered to play the role of donor and QD2 the role of acceptor, it can be concluded that the donor recombines faster than the acceptor, as expected in a transfer process.[5, 41]

This experimental evidence, in connection with the redshift measured both in absorption and steady-state PL spectra, confirms a coupling between the QD1 and the QD2, which is stronger as the interparticle distance decreases.

3.2. Comparison between hetero and homo molecular assemblies

The coupling has been previously demonstrated for homodimers based on two identical QDs linked by dithiols.[5] Here, we consider a more complex system based on QDs of different sizes, physically bounded at a short distance, and we are interested in evaluating how the mutual resonance between the optical transitions of the QDs may affect the coupling. In analogy with the homodimers, characterized by isoenergetic levels, the condition of mutual resonance allows building an energy diagram of the hetero system where the level |1⟩ of the donor and the level |2⟩ of the acceptor are at the same energy (Scheme 2), thus allowing to compare the homo and hetero transfer performances. In a homo system (scheme 2A), the interacting QDs have the same energetic position of all the electronic levels. However, taking into account the intrinsic size dispersion of the QDs ($\sigma\% \sim 7$-$8\%$), the smaller dots in the same batch can act as ''donors'' (QDd) and the larger ones as ''acceptors'' (QDa) and can interact with each other, resulting in coupling phenomena.

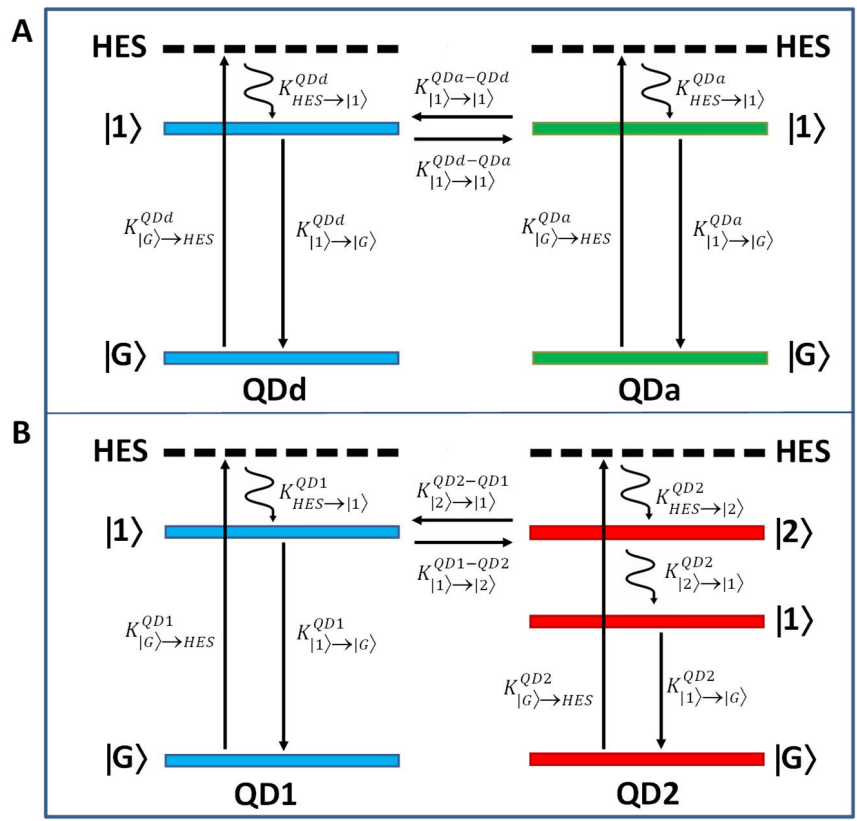

Scheme 2. Energy diagram describing the electronic transitions between two QDs in a homo-assembly (same size, A) and in a hetero-assembly (different size with resonant transition, B). The possible transitions are labelled with the corresponding kinetic constants.

After pumping the QDs to high energy states (HES), the photogenerated electrons thermalize on the level $|1\rangle$ of both QDd and QDa that result almost equally populated. Then, the charge carriers can either relax radiatively back to the ground state $|G\rangle$ with kinetic constant $K^{QD\ a/d}_{|1\rangle\rightarrow|G\rangle}$ in a few nanoseconds or more rapidly transfer or delocalize on the $|1\rangle$ level of the adjacent QD with a kinetic constant $K^{QDd/a-QDa/d}_{|1\rangle\rightarrow|1\rangle}$. In the case of hetero-assemblies (scheme 2B), the electronic level $|1\rangle$ of the QD1 is isoenergetic with level $|2\rangle$ of the QD2, i.e. they are in mutual resonance.

When the hetero system is irradiated at $\lambda_{ex}$=485 nm, both QD1 and QD2 absorb the radiation with kinetic constants $K^{QD1/2}_{|G\rangle\rightarrow HES}$, resulting, after the thermalization process, in a certain population of the level $|1\rangle$ of the QD1 and of level $|2\rangle$ of the QD2. The population of the level $|2\rangle$ of the QD2 thermalizes on the level $|1\rangle$ of the same dot with the kinetic constant $K^{QD2}_{|2\rangle\rightarrow|1\rangle}$ in a sub-ps time scale.[49] Then, after thermalization, the level $|1\rangle$ of the QD1 has a larger population of the level $|2\rangle$ of the QD2, regardless of the different extinction coefficients of the two QDs. Since these two levels are almost isoenergetic (the size dispersion must be considered) a direct coupling is possible. In addition, a more efficient and directional charge transfer from QD1 to QD2 is expected, due to the difference in population, while in the case of the homo-assembly, the probability of transfer between the first excited levels of the two QDs is the same.

To experimentally validate this assumption, a comparison of the spectroscopic properties of homo and hetero molecular assemblies has been carried out. The PL spectrum of pDT hetero sample shows a similar redshift compared to QD1 homo-assembly (Figure S7), being in both cases of a few nm. The same results are obtained for homo systems based on QD2.[5]

A faster PL decay is observed in pDT hetero-assembly (Fig. 5A, yellow line) with respect to the analogous decay measured in homo systems (Table S1) at a short interparticle distance.

Moving to hDT (Fig. 5B), an opposite behaviour is observed, with faster recombination for homo with respect to hetero ones. Similar behaviour is obtained for oDT (Figure S8) and comparing pDT and hDT hetero samples with QD2 pDT homo-assembly (Figure S9). These differences between the TR-PL behaviours when varying the alkyl chain length also reflect in the transfer rates. In Fig. 6, the transfer rates $K_{et}$ are plotted as a function of the interparticle distance (i.e. 0.55 nm for the pDT, 0.9 nm for the hDT, and 1.2 nm for the oDT) for both homo and hetero molecular assemblies. In analogy to[5, 50] $K_{et}$ is defined by the equation $K_{et} = 1/\tau_{(coupled)} - 1/\tau_{(isolated)}$, which respectively considers the exciton average lifetimes in the coupled and isolated pT treated QDs, showing an exponential trend against the ligand length. The trend can be fitted by an exponential function providing a slope parameter that accounts for the different speeds of the transfer processes. The numerical values of $K_{et}$ for each sample are reported in table 1, together with the average lifetime $\tau_{AV}$ and the slope for the three different systems. In general, the coupling phenomena are controlled by the distance between the interacting moieties, ranging from charge transfer (Dexter model) to energy transfer (FRET/Förster theory) [19] and two regimes are expected: a contact zone, where the Dexter model applies, and a near field zone, where the FRET process dominates. The limit between the two zones is defined as 0.01b where b=λ/2πn, λ is the wavelength of the donor fluorescence and n is the refractive index of the solution.[20] Such a value has been calculated to be about 0.7 nm for both homo and hetero molecular systems. Then, it is reasonable to suppose that, below such a distance, the coupling process is a pure charge transfer while, for larger distances, energy transfer mechanisms need to be considered. The Dexter and FRET efficiencies have a different dependence from the distance, namely, in the first case, it depends exponentially (e$^{-r}$) with respect to the distance between the QD surfaces, whereas in the case of the Förster mechanism, efficiency is proportional to $1/d^6$, where d is the center-to-center distance.

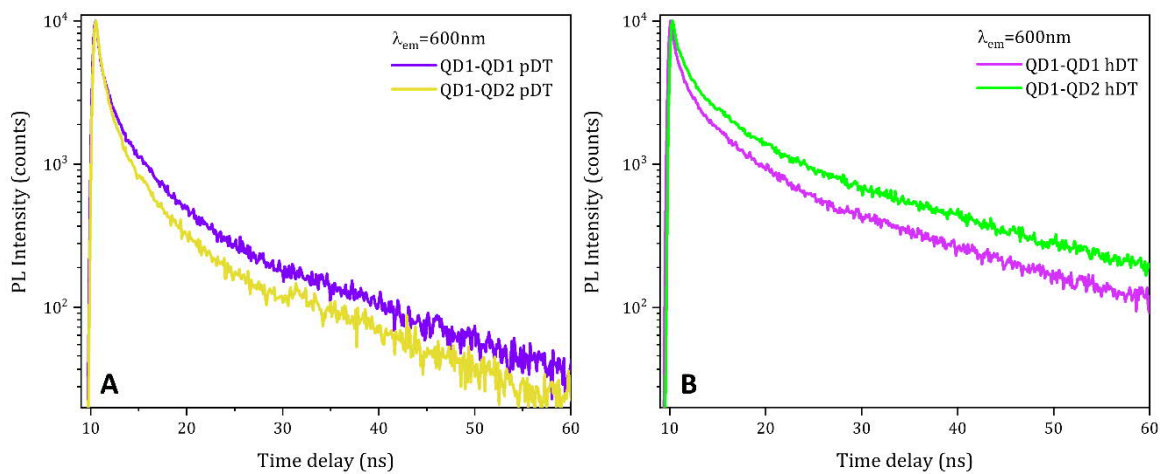

Figure 5. Comparison between TR-PL measurements of A) pDT hetero-assembly and QD1 pDT homo-assembly and B) hDT hetero-assembly and QD1 hDT homo-assembly. $\lambda_{ex}$=485 nm

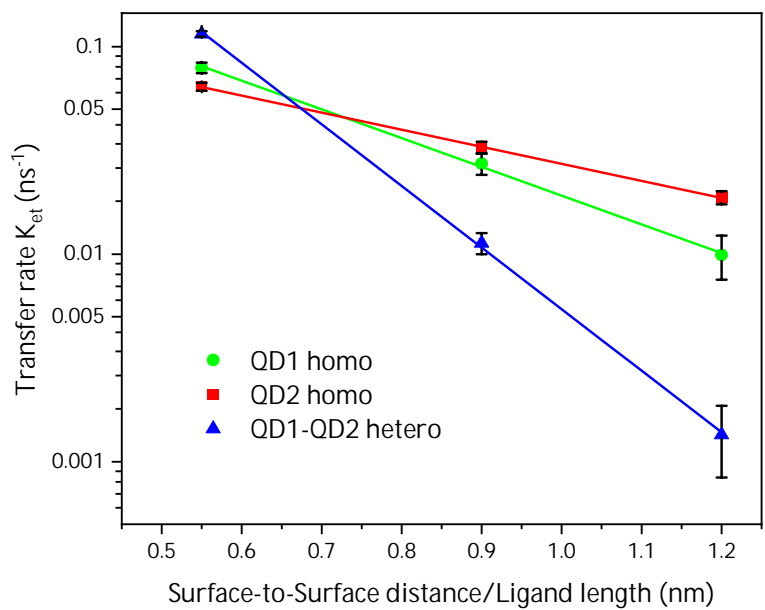

Figure 6. Logarithmic plot of transfer rates calculated from exciton lifetimes, versus ligand length for both the series of homo and the hetero molecular assemblies, fabricated starting from QD1 and QD2. The fitting curves and the error bars are also reported.

| Ligand | Hetero | | QD1 homo | | QD2 homo | |
|---|---|---|---|---|---|---|
| | $\tau_{AV}$ (ns) | $K_{et}$ (ns$^{-1}$) | $\tau_{AV}$ (ns) | $K_{et}$ (ns$^{-1}$) | $\tau_{AV}$ (ns) | $K_{et}$ (ns$^{-1}$) |
| pDT | 6.6 | 0.12 | 8.6 | 0.08 | 10.0 | 0.06 |
| hDT | 18.1 | 0.01 | 13.3 | 0.03 | 14.6 | 0.03 |
| oDT | 22.5 | 0.001 | 21.2 | 0.01 | 16.0 | 0.02 |
| Slope | -2.97 | | -1.38 | | -0.82 | |

Table 1. Decay average lifetime $\tau_{AV}$ and transfer rate $K_{et}$ for the hetero and analogous homo molecular assemblies.

In Fig. 6, at very short distances (when pDT is used as linker) and where the charge transfer process is expected to be dominant, $K_{et}$ for hetero-assemblies is significantly higher than the $K_{et}$ for homo ones (Table 1).

Moreover, at short distances, the homo-assembly based on smaller QDs transfers more efficiently with respect to the large one. Increasing the distance (hDT), the FRET becomes relevant, reflecting in an opposite outcome in the $K_{et}$, resulting faster in both homo-assemblies as compared to the hetero ones. Also, the larger QDs have larger $K_{et}$ than the smaller ones. This behaviour is strongly enhanced when the QDs are linked with oDT (1.2 nm). In particular, the transfer rate of hetero systems increases by two orders of magnitude when reducing the interparticle distance from 1.2 nm (oDT) to 0.55 nm (pDT), while a less marked change is obtained in the case of homo samples, as highlighted in Fig. 6.

These results can be rationalized supposing that, for very short ligand length, the different size of the QDs and probably the mutual resonance of the energy levels facilitate the charge transfer (or wavefunction delocalization) from the donor to the acceptor. In the case of QDs of comparable size, being the delocalization larger for smaller QDs,[5, 51] the charge transfer results more effective for QD1 molecular homo-assembly. At increased interparticle distances, the dipole-dipole interactions start to become relevant, resulting more effective when isoenergetic excitonic levels are coupled. Since in hetero-assembly the first excited levels of QD1 and QD2 have different energies and, accordingly, different associated dipole moments, the hetero sample system shows a reduced coupling in the near field zone where the FRET dominates.[20]

In addition, the FRET efficiency is strictly related to the overlap integral between the emission of the donor and the absorption of the acceptor and then to the molar extinction coefficient of the QDs (higher for larger CdSe QDs). Then, a higher value of $K_{et}$ is obtained for QD2 homo-assemblies. Summarizing, the type of dominant coupling phenomenon is dictated by the interparticle distance between the QDs, ranging from resonant energy transfer (FRET) due to the dipole-dipole interactions at relatively large separation to the overlap of the delocalized wavefunctions of the QDs at very short distances.[20, 21] However, while FRET phenomena are more efficient in molecular homo-assemblies, a high coupling due to wavefunction delocalization is obtained for hetero-assemblies, thanks to the cascade-like energy diagram.

4. Conclusion

A fabrication strategy for the preparation of molecular hetero-assemblies has been successfully translated from an established procedure for homodimers. Molecular QDs assemblies with well-defined interparticle spacing ranging from 1.2 nm down to sub-nm have been achieved by using dithiols with different chain lengths. Hetero systems composed of different sizes QDs, properly designed to have their excitonic energy levels in mutual resonance, have been spectroscopically inspected in solution and compared with the homologs homo-assemblies. The investigation highlights that the QD coupling is governed by the interparticle distance and that QD size and energetic levels significantly affect the transfer rate efficiency. The findings evidence the potential of the two sized system in terms of transfer rate for applications requiring a high charge transfer/wavefunction delocalization, taking advantage of the cascade-like energetic levels in the contact zone. On the other hand, FRET architectures would strongly benefit from the use of homo-assemblies that can couple more efficiently by means of the energy transfer mechanism, thanks to quasi iso-energetic electronic levels. These considerations can be useful to design and engineer more efficiently optoelectronic devices based on QDs assemblies, giving a boost to these applications in the photonic era.

Electronic Supplementary Information (ESI)

Absorbance and PL spectra of QD1, QD2 and mixed QD1-QD2 in solution and TEM characterization; TR-PL decay profiles of QD1 and QD2; TEM micrograph of large aggregates and organic impurities collected at the bottom layer of the centrifuge tube; Crystallographic details of the two QDs; HRTEM image and crystallographic model of oDT hetero-assembly; Average lifetimes $\tau_{AV}\pm\sigma$ (ns) of the mixed QD1-QD2, QD1 and QD2 and relative molecular assemblies fabricated using hDT and pDT as bifunctional linkers; Absorption spectra of the mixed QD1-QD2, treated with pT and with pDT; Normalized PL spectra of the QD1 pDT homo-assembly and mixed pDT hetero-assembly; TR-PL measurements of oDT hetero-assemblies and QD1 oDT homo-assemblies; TR-PL measurements of pDT and hDT hetero-assemblies and QD2 pDT and hDT homo-assemblies; FRET theory.

Author Contributions

Conceptualization: C.N.D., M.S.; methodology: E.F., C.N.D., M.S.; validation: C.N.D., A.P., R.B., L.D., M.S., C.I., R.T., A.A; formal analysis: C.N.D., L.D., R.B., R.T., C.I.; investigation: C.N.D., L.D., R.B., M.S., E.F.; resources: M.S., M.L.C., A.A.; data curation: C.N.D., E.F., C.G., R.B., A.P., R.T.; writing-original draft preparation: C.N.D., M.S., L.D.; writing-review and editing: C.N.D., C.G., E.F., A. P., M.L.C., R.T., M.S.; visualization: C.N.D., A.P., M.L.C., C.I., E.F., R.T., C.G., A.A.; supervision: M.S.; funding acquisition: M.S., M.L.C.

Conflicts of interest

There are no conflicts to declare.

Acknowledgements

This work is financially supported by the European H2020 FET project COPAC (Contract agreement n.766563). The project MIUR PRIN 2015 n. 2015XBZ5YA is also acknowledged.

References

[1] M. Coden, P. De Checchi, B. Fresch, Spectral Shift, Electronic Coupling and Exciton Delocalization in Nanocrystal Dimers: Insights from All-Atom Electronic Structure Computations, Nanoscale, (2020), *12*, 18124-18136, https://doi.org/10.1039/d0nr05601d,
[2] H. Gattuso, R. D. Levine, F. Remacle, Massively Parallel Classical Logic Via Coherent Dynamics of an Ensemble of Quantum Systems with Dispersion in Size, Proc. Natl. Acad. Sci., (2020), *117*, 21022-21030, https://doi.org/10.1073/pnas.2008170117,
[3] M. Yamauchi, S. Masuo, Self-Assembly of Semiconductor Quantum Dots using Organic Templates, Chem. Eur. J., (2020), *26*, 7176-7184, https://doi.org/10.1002/chem.201905807,
[4] H. Gattuso, B. Fresch, R. D. Levine, F. Remacle, Coherent Exciton Dynamics in Ensembles of Size-Dispersed CdSe Quantum Dot Dimers Probed via Ultrafast Spectroscopy: a Quantum Computational Study, Appl. Sci., (2020), *10*, 1328, https://doi.org/10.3390/app10041328,
[5] C. N. Dibenedetto, E. Fanizza, R. Brescia, Y. Kolodny, S. Remennik, A. Panniello, N. Depalo, S. Yochelis, R. Comparelli, A. Agostiano, M. L. Curri, Y. Paltiel, M. Striccoli, Coupling Effects in QD Dimers at Sub-

| | Nanometer Interparticle Distance, Nano Res., (2020), *13*, 1071-1080, https://doi.org/10.1007/s12274-020-2747-3,
[6] Y. Kolodny, S. Fererra, V. Borin, S. Yochelis, C. N. Dibenedetto, M. Mor, J. Dehnel, S. Remmenik, E. Fanizza, M. Striccoli, I. Schapiro, E. Lifshitz, Y. Paltiel, Tuning Quantum Dots Coupling Using Organic Linkers with Different Vibrational Modes, J. Phys. Chem. C, (2020), *124*, 16159-16165, https://doi.org/10.1021/acs.jpcc.0c03703,
[7] J. Cui, Y. E. Panfil, S. Koley, D. Shamalia, N. Waiskopf, S. Remennik, I. Popov, M. Oded, U. Banin, Colloidal Quantum Dot Molecules Manifesting Quantum Coupling at Room Temperature, Nat. Commun., (2019), *10*, 5401, https://doi.org/10.1038/s41467-019-13349-1,
[8] E. Collini, H. Gattuso, Y. Kolodny, L. Bolzonello, A. Volpato, H. T. Fridman, S. Yochelis, M. Mor, J. Dehnel, E. Lifshitz, Y. Paltiel, R. D. Levine, F. Remacle, Room-Temperature Inter-Dot Coherent Dynamics in Multilayer Quantum Dot Materials, J. Phys. Chem. C, (2020), *124*, 16222-16231, https://doi.org/10.1021/acs.jpcc.0c05572,
[9] X. Xu, S. Stöttinger, G. Battagliarin, G. Hinze, E. Mugnaioli, C. Li, K. Müllen, T. Basché, Assembly and Separation of Semiconductor Quantum Dot Dimers and Trimers, J. Am. Chem. Soc., (2011), *133*, 18062-18065, https://doi.org/10.1021/ja2077284,
[10] B. Reznychenko, E. Mazer, M. Coden, E. Collini, C. N. DiBenedetto, A. Donval, B. Fresch, H. Gattuso, N. Gross, Y. Paltiel, F. Remacle, M. Striccoli, An n-Bit Adder Realized via Coherent Optical Parallel Computing, IEEE Int. Conf. Reb. Comp. (ICRC), (2019), 1-7, https://doi.org/10.1109/icrc.2019.8914703,
[11] A. Bordoloi, V. Zannier, L. Sorba, C. Schönenberger, A. Baumgartner, A Double Quantum Dot Spin Valve, Commun. Phys., (2020), *3*, 135, https://doi.org/10.1038/s42005-020-00405-2,
[12] W. Ma, L. Xu, L. Wang, C. Xu, H. Kuang, Chirality-Based Biosensors, Adv. Funct. Mater., (2019), *29*, 1805512, https://doi.org/10.1002/adfm.201805512,
[13] A. O. Govorov, Y. K. Gun'ko, J. M. Slocik, V. A. Gérard, Z. Fan, R. R. Naik, Chiral nanoparticle assemblies: circular dichroism, plasmonic interactions, and exciton effects, J. Mater. Chem., (2011), *21*, 16806-16818, https://doi.org/10.1039/c1jm12345a,
[14] G. S. Selopal, H. Zhao, Z. M. Wang, F. Rosei, Core/Shell Quantum Dots Solar Cells, Adv. Funct. Mater., (2020), *30*, 1908762, https://doi.org/10.1002/adfm.201908762,
[15] C. Wang, D. Barba, G. S. Selopal, H. Zhao, J. Liu, H. Zhang, S. Sun, F. Rosei, Enhanced Photocurrent Generation in Proton-Irradiated "Giant" CdSe/CdS Core/Shell Quantum Dots, Adv. Funct. Mater., (2019), *29*, 1904501, https://doi.org/10.1002/adfm.201904501,
[16] N. Peer, I. Dujovne, S. Yochelis, Y. Paltiel, Nanoscale Charge Separation Using Chiral Molecules, ACS Photonics, (2015), *2*, 1476-1481, https://doi.org/10.1021/acsphotonics.5b00343,
[17] J. Liu, H. Zhang, F. Navarro-Pardo, G. S. Selopal, S. Sun, Z. M. Wang, H. Zhao, F. Rosei, Hybrid surface passivation of PbS/CdS quantum dots for efficient photoelectrochemical hydrogen generation, Appl. Surf. Sci., (2020), *530*, 147252, https://doi.org/10.1016/j.apsusc.2020.147252,
[18] D. L. Dexter, A Theory of Sensitized Luminescence in Solids, J. Chem. Phys., (1953), *21*, 836-850, https://doi.org/10.1063/1.1699044,
[19] T. Förster, Zwischenmolekulare Energiewanderung und Fluoreszenz, Ann. Phys., (1948), *437*, 55-75, https://doi.org/10.1002/andp.19484370105,
[20] B. W. van der Meer, *Förster Theory* in *FRET – Förster Resonance Energy Transfer* (Eds.: I. Medintz, N. Hildebrandt), 2013, pp. 23-62, https://doi.org/10.1002/9783527656028.ch03.
[21] P. Moroz, Z. Jin, Y. Sugiyama, D. A. Lara, N. Razgoniaeva, M. Yang, N. Kholmicheva, D. Khon, H. Mattoussi, M. Zamkov, Competition of Charge and Energy Transfer Processes in Donor–Acceptor Fluorescence Pairs: Calibrating the Spectroscopic Ruler, ACS Nano, (2018), *12*, 5657-5665, https://doi.org/10.1021/acsnano.8b01451,
[22] J. B. Hoffman, H. Choi, P. V. Kamat, Size-Dependent Energy Transfer Pathways in CdSe Quantum Dot–Squaraine Light-Harvesting Assemblies: Förster versus Dexter, J. Phys. Chem. C, (2014), *118*, 18453-18461, https://doi.org/10.1021/jp506757a,
[23] T. Avellini, C. Lincheneau, F. Vera, S. Silvi, A. Credi, Hybrids of semiconductor quantum dot and molecular species for photoinduced functions, Coord. Chem. Rev., (2014), *263-264*, 151-160, https://doi.org/10.1016/j.ccr.2013.07.014,

[24] R. D. Harris, S. Bettis Homan, M. Kodaimati, C. He, A. B. Nepomnyashchii, N. K. Swenson, S. Lian, R. Calzada, E. A. Weiss, Electronic Processes within Quantum Dot-Molecule Complexes, Chem. Rev., (2016), *116*, 12865-12919, https://doi.org/10.1021/acs.chemrev.6b00102,

[25] A. Panniello, M. Trapani, M. Cordaro, C. N. Dibenedetto, R. Tommasi, C. Ingrosso, E. Fanizza, R. Grisorio, E. Collini, A. Agostiano, M. L. Curri, M. A. Castriciano, M. Striccoli, High-Efficiency FRET Processes in BODIPY-Functionalized Quantum Dot Architectures, Chem. Eur. J., (2021), *27*, 2371-2380, https://doi.org/10.1002/chem.202003574,

[26] P. D. Cunningham, A. Khachatrian, S. Buckhout-White, J. R. Deschamps, E. R. Goldman, I. L. Medintz, J. S. Melinger, Resonance Energy Transfer in DNA Duplexes Labeled with Localized Dyes, J. Phys. Chem. B, (2014), *118*, 14555-14565, https://doi.org/10.1021/jp5065006,

[27] P. D. Cunningham, Y. C. Kim, S. A. Díaz, S. Buckhout-White, D. Mathur, I. L. Medintz, J. S. Melinger, Optical Properties of Vibronically Coupled Cy3 Dimers on DNA Scaffolds, J. Phys. Chem. B, (2018), *122*, 5020-5029, https://doi.org/10.1021/acs.jpcb.8b02134,

[28] K. Zheng, K. Žídek, M. Abdellah, N. Zhu, P. Chábera, N. Lenngren, Q. Chi, T. Pullerits, Directed Energy Transfer in Films of CdSe Quantum Dots: Beyond the Point Dipole Approximation, J. Am. Chem. Soc., (2014), *136*, 6259-6268, https://doi.org/10.1021/ja411127w,

[29] J. B. Hoffman, R. Alam, P. V. Kamat, Why Surface Chemistry Matters for QD–QD Resonance Energy Transfer, ACS Energy Lett., (2017), *2*, 391-396, https://doi.org/10.1021/acsenergylett.6b00717,

[30] D. Kim, S. Tomita, K. Ohshiro, T. Watanabe, T. Sakai, I. Y. Chang, K. Hyeon-Deuk, Evidence of Quantum Resonance in Periodically-Ordered Three-Dimensional Superlattice of CdTe Quantum Dots, Nano Lett., (2015), *15*, 4343-4347, https://doi.org/10.1021/acs.nanolett.5b00335,

[31] H. Sugimoto, K. Furuta, M. Fujii, Controlling Energy Transfer in Silicon Quantum Dot Assemblies Made from All-Inorganic Colloidal Silicon Quantum Dots, J. Phys. Chem. C, (2016), *120*, 24469-24475, https://doi.org/10.1021/acs.jpcc.6b08423,

[32] S. Wu, M. McGuigan, A. L. Tiano, S. S. Wong, J. G. Glimm, A First-Principles Study of CdSe Nanoclusters Capped by Thiol Ligands, arXiv preprint: Materials Science, (2013), https://arxiv.org/abs/1308.4671,

[33] N. C. Anderson, M. P. Hendricks, J. J. Choi, J. S. Owen, Ligand Exchange and the Stoichiometry of Metal Chalcogenide Nanocrystals: Spectroscopic Observation of Facile Metal-Carboxylate Displacement and Binding, J. Am. Chem. Soc., (2013), *135*, 18536-18548, https://doi.org/10.1021/ja4086758,

[34] W. W. Yu, L. Qu, W. Guo, X. Peng, Experimental Determination of the Extinction Coefficient of CdTe, CdSe, and CdS Nanocrystals, Chem. Mater., (2003), *15*, 2854-2860, https://doi.org/10.1021/cm034081k,

[35] J. R. Lakowicz, *Time-Domain Lifetime Measurements* in *Principles of fluorescence spectroscopy*, Springer, New York, 2006, pp. 97-155, https://doi.org/10.1007/978-0-387-46312-4_4.

[36] R. Kilaas, Optimal and near-optimal filters in high-resolution electron microscopy, J. Microsc., (1998), *190*, 45-51, https://doi.org/10.1046/j.1365-2818.1998.3070861.x,

[37] J. C. H. Spence, *High-resolution electron microscopy*, 3rd ed., Oxford University Press, New York, 2003, https://doi.org/10.1093/acprof:oso/9780199668632.001.0001.

[38] C. B. Murray, D. J. Norris, M. G. Bawendi, Synthesis and characterization of nearly monodisperse CdE (E = sulfur, selenium, tellurium) semiconductor nanocrystallites, J. Am. Chem. Soc., (1993), *115*, 8706-8715, https://doi.org/10.1021/ja00072a025,

[39] Y. Yin, A. P. Alivisatos, Colloidal nanocrystal synthesis and the organic-inorganic interface, Nature, (2005), *437*, 664-670, https://doi.org/10.1038/nature04165,

[40] Z. A. Peng, X. Peng, Formation of High-Quality CdTe, CdSe, and CdS Nanocrystals Using CdO as Precursor, J. Am. Chem. Soc., (2001), *123*, 183-184, https://doi.org/10.1021/ja003633m,

[41] K. F. Chou, A. M. Dennis, Förster Resonance Energy Transfer between Quantum Dot Donors and Quantum Dot Acceptors, Sensors (Basel), (2015), *15*, 13288-13325, https://doi.org/10.3390/s150613288,

[42] M. La Rosa, E. H. Payne, A. Credi, Semiconductor Quantum Dots as Components of Photoactive Supramolecular Architectures, ChemistryOpen, (2020), *9*, 200-213, https://doi.org/10.1002/open.201900336,

[43]  C. Tanford, *The hydrophobic effect : formation of micelles and biological membranes*, John Wiley & Sons Inc., New York, 1973,

[44]  P. Li, A. Kumar, J. Ma, Y. Kuang, L. Luo, X. Sun, Density gradient ultracentrifugation for colloidal nanostructures separation and investigation, Science Bulletin, (2018), *63*, 645-662, https://doi.org/10.1016/j.scib.2018.04.014,

[45]  O. C. Ibe, *Chapter 4 - Special Probability Distributions* in *Fundamentals of Applied Probability and Random Processes (Second Edition)* (Ed.: O. C. Ibe), Academic Press, Boston, 2014, pp. 103-158, https://doi.org/10.1016/B978-0-12-800852-2.00004-3.

[46]  T. C. Kippeny, M. J. Bowers, A. D. Dukes, J. R. McBride, R. L. Orndorff, M. D. Garrett, S. J. Rosenthal, Effects of surface passivation on the exciton dynamics of CdSe nanocrystals as observed by ultrafast fluorescence upconversion spectroscopy, J. Chem. Phys., (2008), *128*, 084713, https://doi.org/10.1063/1.2834692,

[47]  S. F. Wuister, C. de Mello Donegá, A. Meijerink, Influence of Thiol Capping on the Exciton Luminescence and Decay Kinetics of CdTe and CdSe Quantum Dots, J. Phys. Chem. B, (2004), *108*, 17393-17397, https://doi.org/10.1021/jp047078c,

[48]  B. Kundu, S. Chakrabarti, A. J. Pal, Redox Levels of Dithiols in II–VI Quantum Dots vis-à-vis Photoluminescence Quenching: Insight from Scanning Tunneling Spectroscopy, Chem. Mater., (2014), *26*, 5506-5513, https://doi.org/10.1021/cm501469f,

[49]  Q. Darugar, C. Landes, S. Link, A. Schill, M. A. El-Sayed, Why is the thermalization of excited electrons in semiconductor nanoparticles so rapid? Studies on CdSe nanoparticles, Chem. Phys. Lett., (2003), *373*, 284-291, https://doi.org/10.1016/S0009-2614(03)00213-6,

[50]  J. J. Choi, J. Luria, B.-R. Hyun, A. C. Bartnik, L. Sun, Y.-F. Lim, J. A. Marohn, F. W. Wise, T. Hanrath, Photogenerated Exciton Dissociation in Highly Coupled Lead Salt Nanocrystal Assemblies, Nano Lett., (2010), *10*, 1805-1811, https://doi.org/10.1021/nl100498e,

[51]  R. Koole, P. Liljeroth, C. de Mello Donegá, D. Vanmaekelbergh, A. Meijerink, Electronic Coupling and Exciton Energy Transfer in CdTe Quantum-Dot Molecules, J. Am. Chem. Soc., (2006), *128*, 10436-10441, https://doi.org/10.1021/ja061608w,

# Coupling in Quantum Dot Molecular Hetero-Assemblies

Carlo Nazareno Dibenedetto[a, b], Elisabetta Fanizza[a, b], Liberato De Caro[c], Rosaria Brescia[d], Annamaria Panniello[b], Raffaele Tommasi[e, b], Chiara Ingrosso[b], Cinzia Giannini[c], Angela Agostiano[a, b], Maria Lucia Curri[a, b], Marinella Striccoli[b, *]

[a.] *Department of Chemistry, University of Bari ''Aldo Moro'', Via Orabona, 4 - 70125 Bari (Italy).*
[b.] *Institute for chemical and physical processes of CNR (IPCF-CNR), Via Orabona, 4 - 70125 Bari (Italy).*
[c.] *Institute of Crystallography of CNR (CNR-IC), Via Amendola, 122/O - 70125 Bari (Italy)*
[d.] *Italian Institute of Technology (IIT), Via Morego 30 - 16163 Genova (Italy).*
[e.] *Basic Medical Sciences, Neuroscience and Sense Organs, University of Bari ''Aldo Moro'', Piazza Giulio Cesare, 11 - 70124 Bari (Italy).*

Additional data

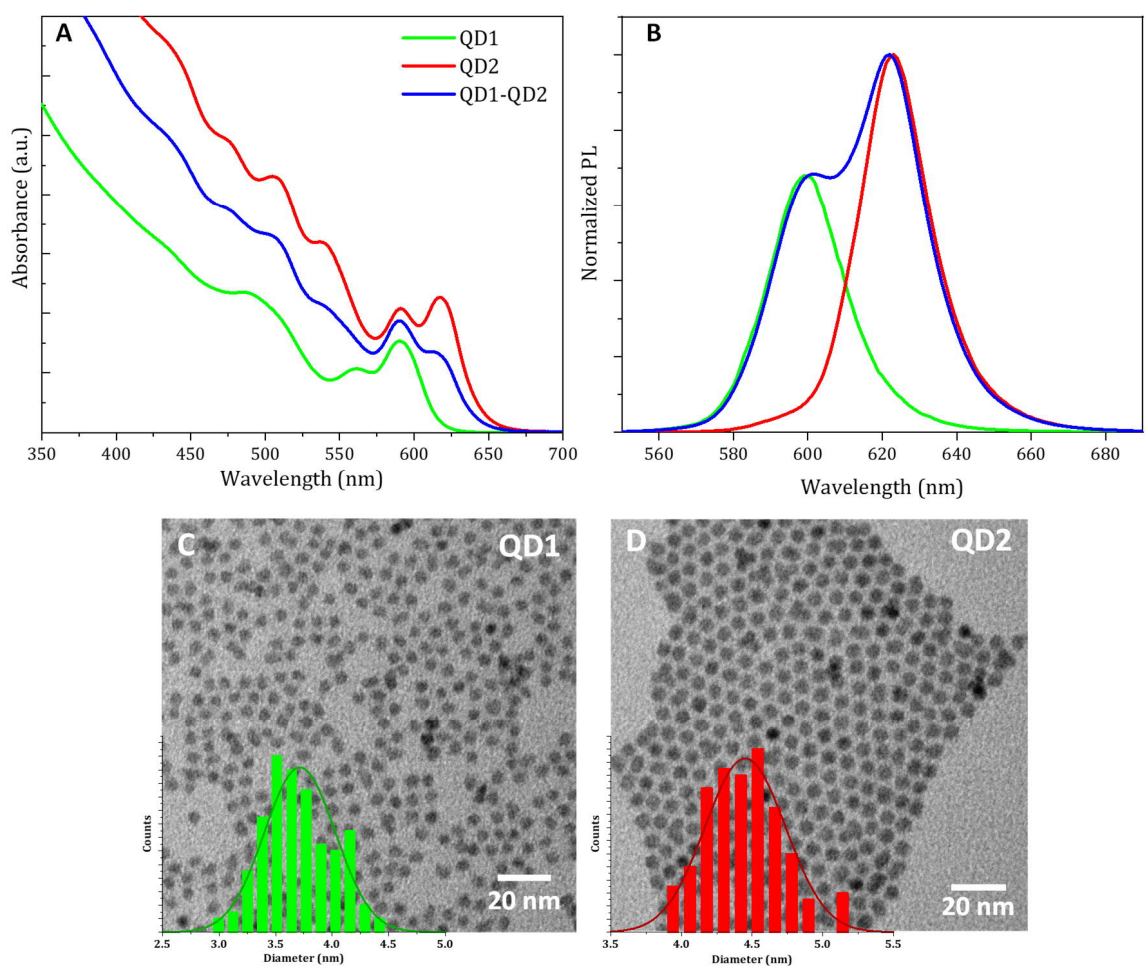

Figure S 1. UV-Vis Absorption (A), normalized PL spectra at $\lambda_{ex}$=485 nm (B), and bright-field TEM images with size distribution histogram (C, D) of the QDs used for the fabrication of molecular hetero-assemblies. The green line refers to QD1, the red line to QD2, and the blue line to the QD1-QD2 1:1 mixed solution. The PL spectra of QD1 and QD2 have been normalized with respect to the maximum emission wavelength of the respective QDs in the mixed solution.

The absorption (S1A) and the PL (S1B) spectra of the solution of mixed QD1 and QD2, both at the same molar concentration of $5 \cdot 10^{-7}$ M are reported in blue. The absorption profile is characterized by peak at 617 nm, mainly ascribed to the first excitonic transition of QD2 and a higher energy peak at 590 nm, more intense, due to the contribution of both the first transition of QD1 and the second excitonic transition of QD2. The normalized PL spectrum of the mixed QD1-QD2 solution is simply given by the superposition of the PL spectra of the single QDs, thus confirming that the nanoparticles do not interact in absence of linkers.

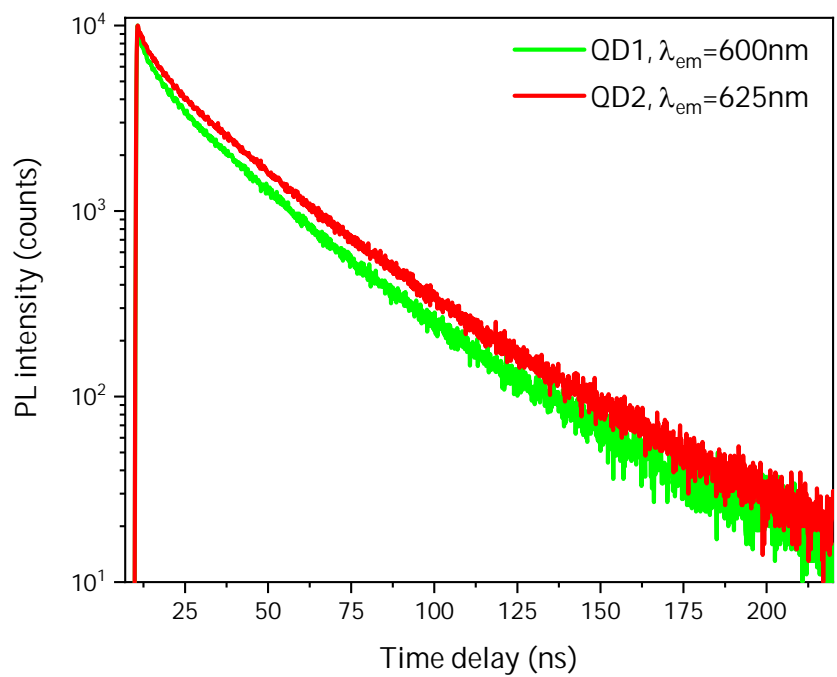

Figure S2. TR-PL decay profiles of QD1 and QD2, used as nano-building block for the fabrication of homo and hetero molecular assemblies. $\lambda_{ex}$=485 nm

In Figure S2 the TR-PL decay profiles of QD1 and QD2 are reported. The calculated $\tau_{AV}$ for these samples are (26.8 ± 0.7) ns and (28.0 ± 0.8) ns, respectively.

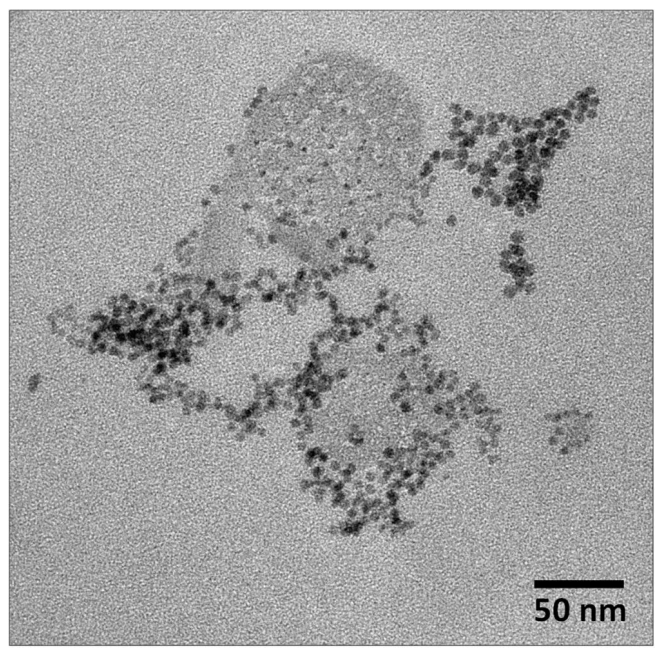

Figure S3. TEM micrograph of the large aggregates and organic impurities collected at the bottom layer of the centrifuge tube in the DGU.

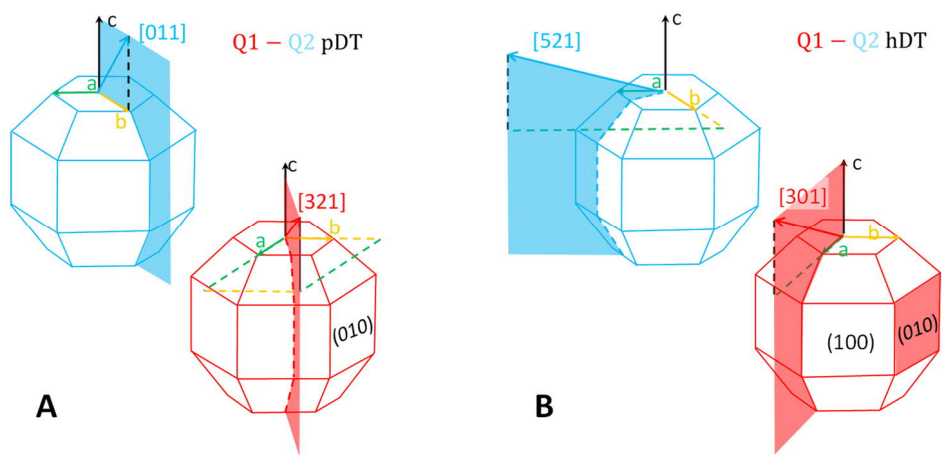

Figure S4. Crystallographic details of the two QDs visualized in the HR-TEM image for pDT (A) and hDT (B) molecular hetero-assemblies. The zone-axis projection of the 3D-model is reported.

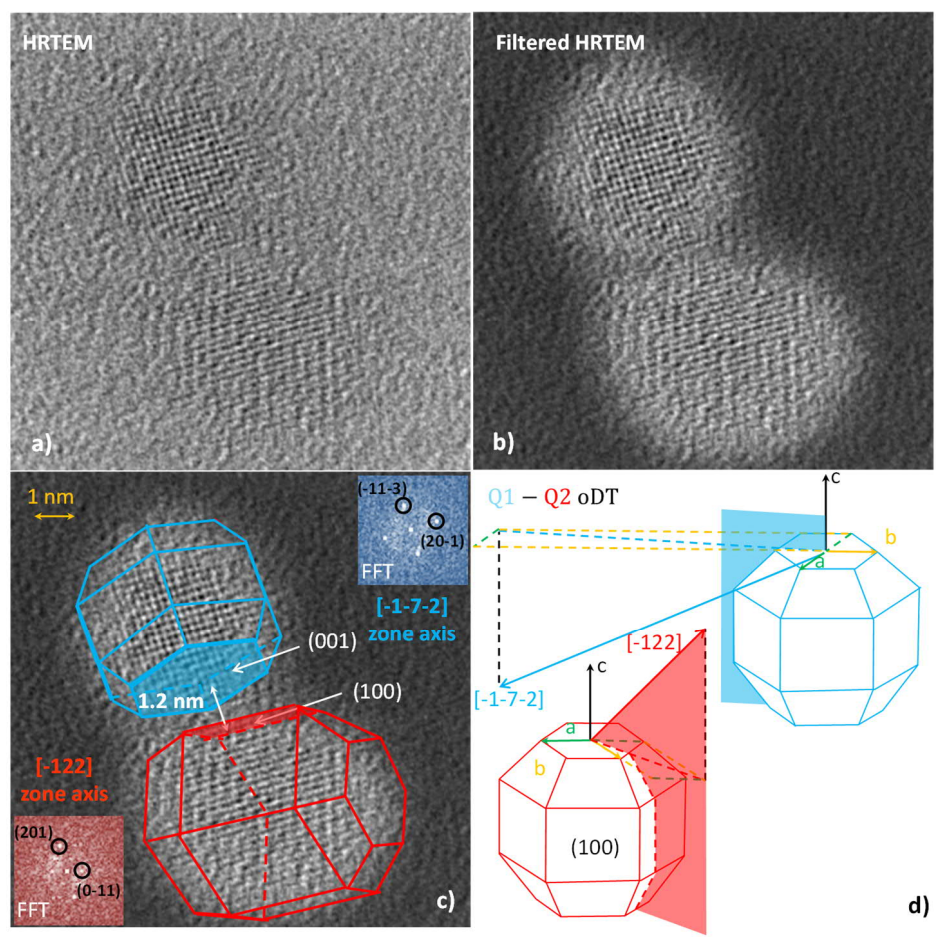

Figure S5. Molecular hetero-assemblies fabricated with oDT. a) Filtered HR-TEM image of the QDs, to minimize high-frequency noise. b) Further filtering of the HR-TEM image to enhance the QDs with respect to the amorphous background. c) 3D-model (frustrated hexagonal prismatic shape) of the two QDs constituting the molecular assembly, seen in projection along the respective zone-axis direction. The insets show the FFTs of the QDs. The nearest surface faces of the QDs are the (100) and the (001). d) Crystallographic details of the two QDs visualized in the HR-TEM image with the zone-axis projection of the 3D-model.

In Figure S5 a segment of length 1.2 nm – the nominal length of oDT chains – between the two QDs is sketched, to give a length reference, to show as the QDs' distance values that can be estimated by the 3D-model reconstruction are in agreement with the nominal values of the linear alkyl chain lengths.

|  | $\lambda_{em}$ =600 nm | $\lambda_{em}$ =625 nm |
|---|---|---|
| **Heterodimers QD1-QD2 averaged lifetimes $\tau_{AV} \pm \sigma$ (ns)** | | |
| QD1-QD2 | 25.2 ± 0.7 | 26.0 ± 0.7 |
| QD1-QD2 pT | 23.6 ± 0.5 | 25.1 ± 0.5 |
| QD1-QD2 hDT | 18.1 ± 0.3 | 21.0 ± 0.5 |
| QD1-QD2 pDT | 6.6 ± 0.1 | 8.6 ± 0.2 |
| **Homodimers QD1 $\tau_{AV} \pm \sigma$ (ns)** | | |
| QD1 | 26.8 ± 0.7 | |
| QD1 pT | 26.1 ± 0.4 | |
| QD1-QD1 hDT | 13.3 ± 0.3 | |
| QD1-QD1 pDT | 8.6 ± 0.2 | |
| **Homodimers QD2 $\tau_{AV} \pm \sigma$ (ns)** | | |
| QD2 | | 28.0 ± 0.8 |
| QD2 pT | | 26.9 ± 0.6 |
| QD2-QD2 hDT | | 14.6 ± 0.4 |
| QD2-QD2 pDT | | 10.0 ± 0.3 |

Table S1. Average lifetimes $\tau_{AV} \pm \sigma$ (ns) of the mixed QD1-QD2, QD1 and QD2 and relative molecular assemblies fabricated using hDT and pDT as bifunctional linkers. The $\tau_{AV}$ are obtained by fitting with multi-exponential functions the TR-PL decay profiles of the different samples and calculating the average lifetime according to ref.[1]

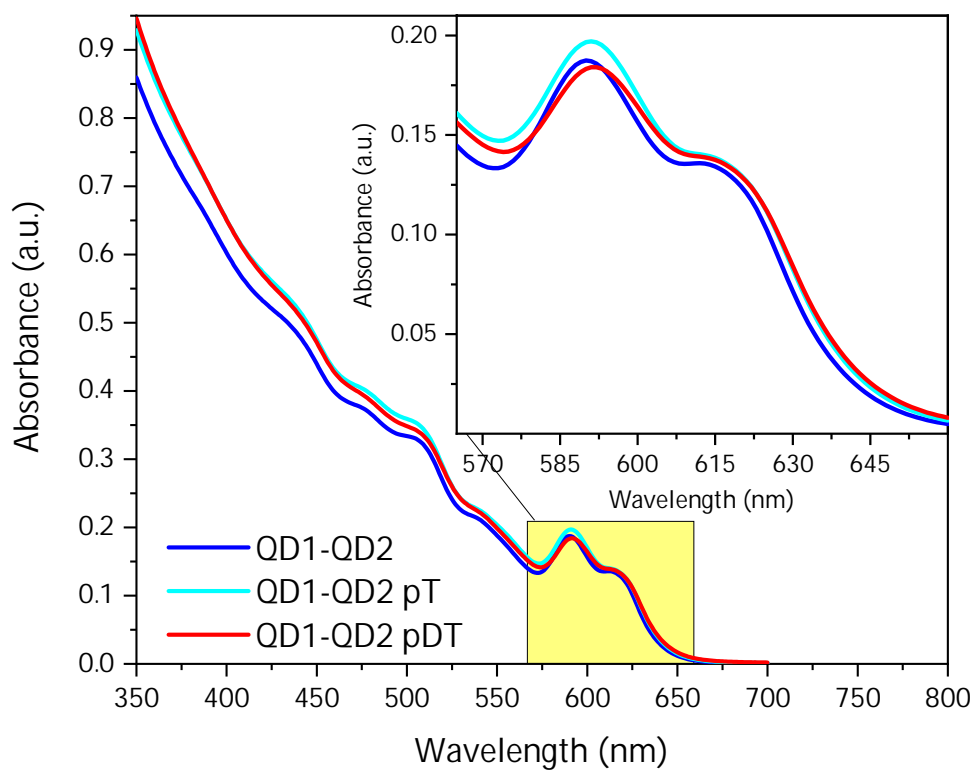

Figure S6. Absorption spectra of the mixed QD1-QD2 (blue line), treated with pT (cyan line) and with pDT (hetero-assembly, red line). In the inset, a magnification of the first excitonic resonance, highlighting the redshift occurring upon treatment with pDT, evidence of the molecular assembly formation.

In Figure S6 the absorption spectra of the mixed QDs, the pT-treated QDs and the pDT molecular hetero-assemblies have been reported. A slight increment at high energy, ascribed to scattering for the samples treated with the alkyl thiols class, and a small red shift for QD1-QD2 sample treated with pDT can be observed. The absorbance of the excitonic peak relative to QD1 in the mix solution with pT results slightly higher than the analogue of the QD1-QD2 solution without pT, while this does not appear to be the case for QD2. Such a difference reflects also in the PL spectrum shown in Figure 3. The origin of this behaviour could be attributed to a slightly higher concentration of the smaller QDs in the mixed solution. The variation in absorbance is really minimal and does not translate into significant changes, in the case of functionalization with dithiols.

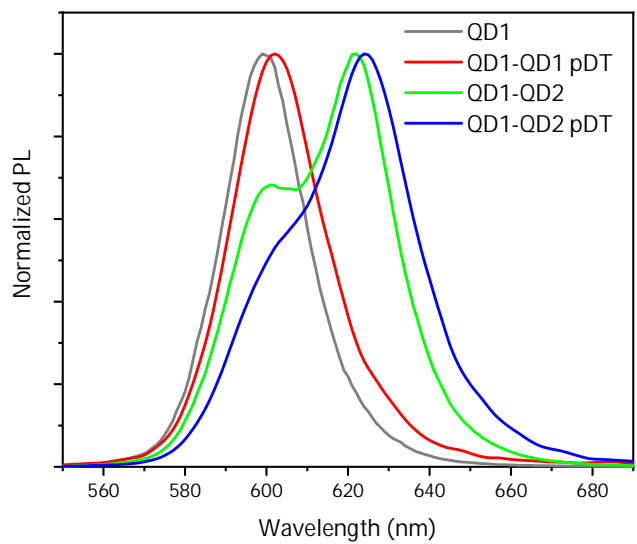

Figure S7. Normalized PL spectra of the QD1 pDT homo-assembly (red line) and pDT hetero-assembly (blue line). For reference, also the emission spectra of the QD1 (grey line) and the mixed QDs (green line) are reported. $\lambda_{ex}$=485 nm

In figure S7 the comparison between the PL spectra of homodimers and heterodimers shows a clear redshift in presence of pDT in both cases with respect to the not functionalized QDs, due to the delocalization and overlap of the wavefunctions in the interspace between the QDs linked by pDT. In addition, for the molecular hetero-assemblies, a further quenching of the peak at 600 nm (QD1) is observed, suggesting that a transfer from QD1 (donor) to QD2 (acceptor) takes place. At this very short interparticle distance (pDT ~ 0.55 nm) a charge transfer is expected, as discussed in the paper.

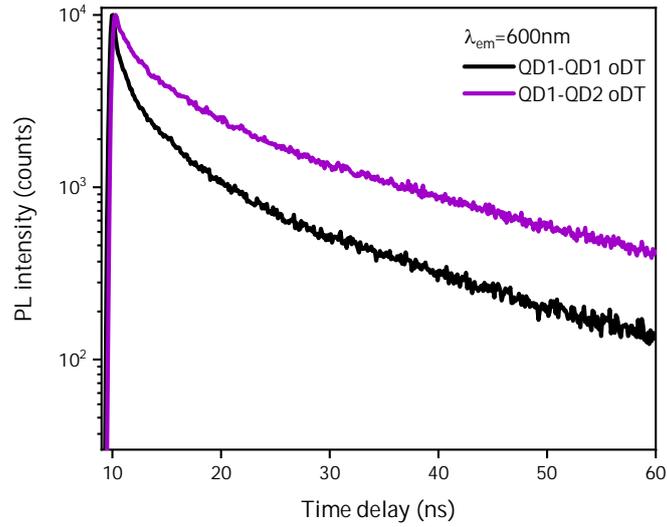

Figure S8. TR-PL measurements of oDT molecular hetero-assembly and QD1 oDT homo-assembly.

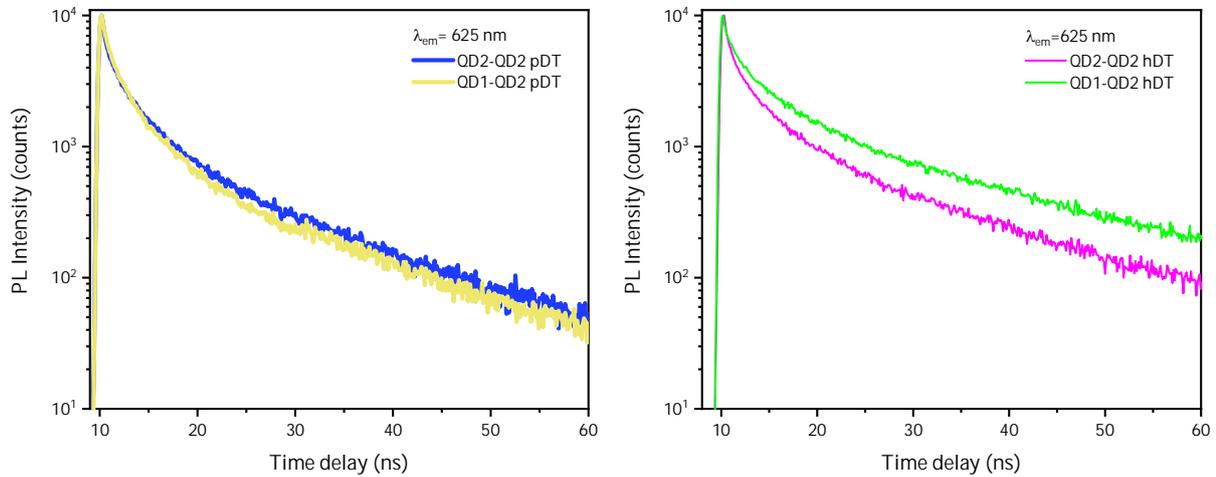

Figure S9. Comparison between TR-PL measurements of pDT hetero-assembly and QD2 pDT homo-assembly (left panel) and hDT hetero-assembly and QD2 hDT homo-assembly (right panel), at $\lambda_{ex}$=485 nm

FRET Theory

The theoretical FRET efficiency[1] between two interacting moieties can be calculated by using the equation $E_{FRET} = \frac{R_0^6}{R_0^6 + r^6}$, where $R_0^6 = \frac{9(ln10)k^2\Phi_D}{128\pi^5 n^4 N_A} J(\lambda)$. $R_0$, the donor–acceptor separation where energy transfer has a 50% probability of occurring, depends from $J(\lambda)$, that represents the normalized overlap integral between the fluorescence of the donor and the absorbance of the acceptor, the QY ($\Phi_D$) of the donor (~24% for both QD1 and QD2), the refractive index of the solvent n (for hexane n=1.37) and $k^2$ is a parameter given by the mutual orientation of the two dipoles of the acceptor and the donor. Experimentally, a FRET efficiency of 28% can be calculated for the hDT molecular hetero-assemblies and 50% for hDT homo ones by using the formula: $E_{FRET}$=1-$\tau_{D-A}$/$\tau_D$, with $\tau_{D-A}$ the lifetime of the donor in the presence of the acceptor and $\tau_D$ the lifetime of the donor.[1]

Lifetime calculation

The PL decays have been fitted by a multi-exponential function, I(t).[2] In the multi-exponential model, the emission intensity is assumed to decay as the sum of individual single exponential decays, as reported in equation S1:

$$I(t) = \sum_i \alpha_i \exp(-t/\tau_i) \tag{S1}$$

where $\tau_i$ are the decay times and $\alpha_i$ represent the amplitudes of the components at t = 0, while the average lifetime $\bar{\tau}$ is given by the equation S2

$$\bar{\tau} = \frac{\sum_i \alpha_i \tau_i^2}{\sum_i \alpha_i \tau_i} \tag{S2}$$

[1]    J. R. Lakowicz, Principles of fluorescence spectroscopy, Springer, New York, 2006, Cap.13 Energy Transfer  pp. 443-475, https://doi.org/10.1007/978-0-387-46312-4_13.
[2]    J. R. Lakowicz, Principles of fluorescence spectroscopy, Springer, New York, 2006, Cap.4 Time-Domain Lifetime Measurements  pp. 141-142, https://doi.org/10.1007/978-0-387-46312-4_4.